\documentclass{article}
\begin{document}

\begin{center}
\centerline{\large \bf Comment on "Cooling atoms with a moving one-way 
barrier"} 
\end{center}

\vspace{3 pt}
\centerline{\sl V.A. Kuz'menko\footnote{Electronic 
address: kuzmenko@triniti.ru}}

\vspace{5 pt}

\centerline{\small \it Troitsk Institute for Innovation and Fusion 
Research,}
\centerline{\small \it Troitsk, Moscow region, 142190, Russian 
Federation.}

\vspace{5 pt}

\begin{abstract}

E.A. Schoene et al., in arXiv:1012.3207 [Phys.Rev.A {\bf82}, 023419 (2010)] 
give incorrect interpretation of their experimental results. Really we 
observe a temporary capture and additional cooling of atoms in the presence 
of the orthogonal laser beam. Any strong potential barriers are absent here. 
It may be easily experimentally tested.
       
\vspace{5 pt}
\end{abstract}

\vspace{12 pt}

The authors present in [1] the results of their experimental study of 
cooling of atoms with a moving one-way barrier. This work continues early 
published experimental study of so-called optical one-way barrier for 
neutral atoms [2, 3]. The base of this works is a widespread now opinion 
about efficient polarization interaction between neutral cold atoms and 
intense nonresonant laser light [4]. It is proposed that a blue-detuned 
light creates a repulsive potential for atoms, and, oppositely, a 
red-detuned light produces an attractive potential. The authors believe 
that just the same property of the barrier beam defines the observed motion 
of the cold atoms in the published experiments [1-3]. 
	
However, we believe that this is a wrong explanation. The experimental 
results in [1] show only quite usual temporary capture and some cooling of 
atoms in the presence of additional orthogonal laser beam similar, for 
example, to [5]. As a whole, in the discussed experiments with the "barrier" 
beams [2, 3] the motion of atoms is determined not by any potential 
barriers, but by the asymmetry of Raman transitions in the strong main trap 
beam [6]. And the "barrier" beam only denotes the initial position of the 
atoms in the space.
	
So, the mechanism of "reflection" of the atoms from the "barrier" beam may 
be explained in a following way. The repumping beam leads to a spontaneous 
emission of a photon by excited atoms, which erases any its memory about 
the initial state. Then in the main "barrier" beam the Raman transition 
takes place and writes in the atom's memory the information about its new 
initial position in the space. At the next step the asymmetrical Raman 
transition in the field of the main trap beam occurs in such way that the 
photon's recoil moment is always directed against the atom's motion toward 
its new marked initial point in space. It looks like as a "reflection" of 
the atoms from the "barrier" beam.
	
The described above process is also the main mechanism of cold atoms trapping 
and cooling in a crossed laser beams. The main base of this phenomenon is a 
fundamental property of quantum physics: inequality of forward and reversed 
transitions or time reversal noninvariance [7]. 
	
The possible role of the motion of the "barrier" in [1] for additional atom 
cooling is not an obvious conclusion. The additional atom cooling should 
takes place also in the fixed orthogonal laser beam, as it happened, 
for example, in [5]. 
	
So, we have two alternative physical explanations of the discussed "barrier" 
phenomenon: the existence of a potential barriers and the asymmetry of Raman 
transitions. The simple experimental test may be used here. The arrangement 
of experiments in [1-3] should be changed from horizontal to vertical 
position of the main trap beam. If it will be switched of, the atoms will 
freely fall down under action of gravity. And we can see two possible cases:

1 - the atoms will freely pass through the "barrier" beams,

2 - the atoms will be reflected from the barrier beams. 

We believe that the first variant will be observed because of the intensity 
of the main "barrier" beam is insufficiently high for itself capture and 
trapping of the atoms, as, for example, in [8]. 
	
It is pity that the authors of [1-3] waste their efforts to the study of the 
phantom instead of to study the really important physical problem of 
inequality of forward and reversed processes in quantum physics. This is not 
very difficult task. Our scientists should only at last wake up and stop to 
trust in the old myth that lows in quantum physics are invariant under time 
reversal [9, 10].

\end{document}